\pdfoutput=1
\documentclass[11pt,a4paper]{article}

\usepackage[T1]{fontenc}
\usepackage{mathpazo}
\usepackage[margin=25mm]{geometry}
\usepackage{amsmath}
\usepackage{microtype}
\usepackage{xcolor}
\usepackage{titlesec}
\usepackage{array,booktabs,tabularx}
\usepackage[font=small,labelfont=bf]{caption}
\usepackage{enumitem}
\usepackage{tcolorbox}
\usepackage{graphicx}
\usepackage[hidelinks]{hyperref}
\definecolor{accent}{RGB}{31,58,95}

\titleformat{\section}{\large\bfseries\color{accent}}{\thesection}{0.8em}{}
\titleformat{\subsection}{\normalsize\bfseries\color{accent}}{\thesubsection}{0.8em}{}
\titlespacing*{\section}{0pt}{14pt}{6pt}
\titlespacing*{\subsection}{0pt}{10pt}{4pt}

\setlist{itemsep=2pt, topsep=2pt}

\newtcolorbox{example}[1]{%
  colback=accent!4!white, colframe=accent!40!white,
  colbacktitle=accent!10!white, coltitle=accent,
  fonttitle=\small\bfseries, title={#1},
  boxrule=0.5pt, left=6pt, right=6pt, top=4pt, bottom=4pt}

\newtcolorbox{note}{%
  colback=black!3, colframe=black!25,
  boxrule=0.4pt, left=6pt, right=6pt, top=4pt, bottom=4pt, fontupper=\small}

\newcommand{\pt}{\mathrm{pt}}
\DeclareMathOperator{\median}{median}

\title{\color{accent}\bfseries A joint error metric for scoring CFD against\\
  wind tunnel data, scaled by the robust spread\\
  of the participants (AutoCFD5 Case 2)}
\author{H. Hetmann, C. Mockett\\[2pt]{\small Upstream CFD GmbH, Berlin}}
\date{October 2026}

\begin{document}

\maketitle

\begin{abstract}
\noindent Case 2 of the 5th Automotive CFD Prediction Workshop
(AutoCFD5)\footnote{\url{https://autocfd.org}} compares the participants' CFD results with wind
tunnel measurements of surface
pressure, velocity profiles and force coefficients. To place each run on a diagram of error against
computational cost, these quantities must be combined into one number. This note defines that
number, the \emph{joint error}. Following the practice of interlaboratory proficiency testing
(ISO 13528), every deviation from the experiment is expressed in units of the spread of the
participants' results at that point, estimated robustly with the standard's Algorithm~A. The
point-wise spreads are pooled into one scale per component, so that every measurement point
carries equal weight within its component. Configurations 2a and 2b share these scales and are
compared only on the measurements both have, so that their errors can be shown on one diagram. The
joint error is a fixed weighted sum of five
component errors. The scales are computed once from all AutoCFD5 submissions received by
28~October 2026 and then frozen, so that scores are reproducible and comparable across workshops.
\end{abstract}

\section{Motivation}

\subsection{The task}

Case 2 is scored in three \emph{variants}: the two configurations 2a and 2b, and their difference
DD${}={}$2b${}-{}$2a. The DD variant tests whether a run captures the effect of the change from one
configuration to the other, rather than each configuration on its own. Every variant is scored
separately, so each run receives three joint errors. For each variant, the wind tunnel provides
\begin{itemize}
  \item the pressure coefficient $C_p$ at about 200 surface probes,
  \item the normalised velocity magnitude $V/V_\infty$ at 400 to 630 points on six (2b) or eleven
    (2a) of 16 measurement lines, and
  \item the drag, lift and pitching moment coefficients $C_D$, $C_L$ and $C_M$.
\end{itemize}
The workshop compares codes and set-ups by plotting an error measure against a cost measure. Runs
that no other run beats on both axes form the Pareto front. This requires a single error value
per run and variant, lower being better.

We call each of these five physical quantities a \emph{component}: the pressure coefficient
$C_p$, the velocity $V/V_\infty$, and the force and moment coefficients $C_D$, $C_L$ and $C_M$.
The components differ in kind, in number of points and in magnitude. A $C_p$ deviation of 0.05,
a velocity deviation of 0.05 and a drag deviation of 0.005 cannot be added as they stand. Each
component needs a scale, and the choice of scales decides how much each component counts.

\subsection{Choosing the scale}
\label{sec:scale}

Three sources of scale are available.

\textbf{Fixed tolerances.} One could divide each deviation by a tolerance set by judgement, for
example 0.025 for $C_D$. In a weighted sum $\sum_k w_k\,|\Delta C_k|/t_k$, however, the effective
weight of a component is $w_k/t_k$: halving a tolerance doubles the component's weight. The
tolerances become hidden weights that are hard to justify.

\textbf{The spread of the experimental data.} Profiles can be scaled by the variation of the
measured values along the profile, as in the coefficient of determination $R^2$. This measures
how much a profile varies, not how hard it is to predict, and it has no equivalent for a single
force coefficient.

\textbf{The spread of the participants' results.} This is the established answer of
\emph{proficiency testing}. When laboratories measure the same item, each result is compared with
an assigned value and the deviation is expressed in units of a standard deviation for proficiency
assessment, $\sigma_\pt$. The requirements for such schemes are set out in ISO/IEC 17043
\cite{iso17043}, and the international standard for their statistics is ISO 13528:2022
\cite{iso13528}. One of its accepted ways of setting $\sigma_\pt$ is the robust spread of the
participants' own results in the same round (clause 8.6). The deviation then becomes a $z$ score
(clause 9.4),
\begin{equation}
  z_i = \frac{x_i - x_{\pt,i}}{\sigma_{\pt,i}},
  \label{eq:z}
\end{equation}
where $x_\pt$ is the assigned value.

The analogy with Case 2 is direct. The CFD runs play the role of the laboratories, and the wind
tunnel result is the assigned value. Taking every scale from the participants' spread gives all
components the same unit: a component error of 1 means ``one typical participant spread away from
the experiment'', whether the component is a pressure distribution or a drag coefficient. The
scales are derived by one standard method, and the main judgement left in the metric is the set
of weights (Section~\ref{sec:errors}).

\section{Method}

\subsection{Deviations from the experiment}

For every run and variant, the deviation from the experiment is computed at every measurement
point $i$ for which an experimental value exists:
\begin{equation}
  d_i = x_{\mathrm{CFD},i} - x_{\mathrm{exp},i}.
\end{equation}
\begin{itemize}
  \item \textbf{Force coefficients.} $C_D$ and $C_L$ are taken as submitted. The pitching moment
    coefficient about the wheelbase centre, referred to the wheelbase, follows from the axle lift
    coefficients as $C_M = (C_{L,f} - C_{L,r})/2$.
  \item \textbf{Surface pressure.} Probes are matched by probe number. Because participants and
    the wind tunnel use different reference pressures, each CFD distribution is first shifted by
    a constant so that it matches the experiment at probe 107. Probe 107 itself is therefore not
    compared.
  \item \textbf{Velocity.} Participants sample each line as they choose. Each CFD profile is
    interpolated linearly onto the experimental points. If a CFD line starts late or ends early,
    its first or last value is held over at most two experimental points. Points further out are
    treated as missing.
  \item \textbf{Variant DD.} The deltas $x_{\mathrm{2b}} - x_{\mathrm{2a}}$ are formed for each
    run and for the experiment, and are then treated exactly like the values of a configuration.
\end{itemize}

\subsection{Participant spread at each point}

At every measurement point, $\sigma_{\pt,i}$ is the robust standard deviation of all runs'
results at that point. Since the experimental value is a constant at each point, this equals the
spread of the deviations $d_i$.

The robust standard deviation is computed with \emph{Algorithm A} of ISO 13528:2022, Annex C,
which originates in ISO 5725-5 \cite{iso5725}. For the $p$ results $x_1, \dots, x_p$ at a point:

\begin{enumerate}[itemsep=4pt]
  \item \textbf{Start from medians.} Set $x^* = \median(x_j)$ and
    $s^* = 1.483 \cdot \median |x_j - x^*|$.

    \textit{The factor 1.483.} The median absolute deviation of normally distributed data is
    $0.6745\,\sigma$. The factor $1.483 = 1/0.6745$ turns it into an estimate of $\sigma$.
  \item \textbf{Clip.} With $\delta = 1.5\,s^*$, replace every value below $x^* - \delta$ by
    $x^* - \delta$ and every value above $x^* + \delta$ by $x^* + \delta$.
  \item \textbf{Recompute.} Set $x^*$ to the mean and $s^*$ to 1.134 times the sample standard
    deviation of the clipped values.

    \textit{The factor 1.134.} Clipping normally distributed data at $\pm 1.5\,\sigma$ reduces
    their standard deviation by a factor of about 0.88. The factor $1.134 \approx 1/0.88$
    restores it.
  \item \textbf{Repeat} steps 2 and 3 until $x^*$ and $s^*$ no longer change. Then
    $\sigma_{\pt,i} = s^*$.
\end{enumerate}

For normally distributed results without outliers, Algorithm A therefore returns the ordinary
standard deviation.

Clipping in this way is called \emph{winsorising}, after the biostatistician Charles P. Winsor. A
distant result is not discarded: it keeps its vote, but its influence is limited to 1.5 spreads
from the centre. Because the clipping limit depends on the spread and the spread depends on the
clipping, the result is defined implicitly, as a fixed point; the iteration is simply a reliable
way to find it. Statisticians will recognise a Huber-type M-estimator \cite{huber}.

\begin{example}{Example: one wild result among eight}
Eight runs give the values 9.8, 9.9, 9.9, 10.0, 10.1, 10.1, 10.2 and 15.0.
\begin{itemize}[leftmargin=12pt]
  \item \textit{Start.} The median is $x^* = 10.05$. The absolute deviations from it have a median
    of 0.15, so $s^* = 1.483 \times 0.15 = 0.22$.
  \item \textit{First clip.} The band is $10.05 \pm 1.5 \times 0.22$, i.e.\ 9.72 to 10.38. The
    value 15.0 is pulled in to 10.38; the others are unchanged.
  \item \textit{Converged.} $x^* = 10.04$ and $\sigma_\pt = 0.20$.
\end{itemize}
For comparison, the ordinary standard deviation of the eight values is 1.77, and that of the
seven agreeing values alone is 0.14. The eighth run still counts, but it no longer inflates the
spread twelvefold.
\end{example}

\textbf{Which runs count.} Every valid run is one result, including several runs from the same
participant with different codes or set-ups. Withdrawn runs and the experiment itself are excluded.

\textbf{Runs that cannot be told apart.} Some participants submitted studies around one set-up:
the same simulation on different hardware, with a different solver tolerance, or with small
geometry details changed. Such runs often give practically the same result. Counted separately,
they would give one set-up as many votes as it has runs and pull the spread towards it. A
participant's runs that cannot be told apart therefore count as one result, their mean. Two runs
$a$ and $b$ cannot be told apart if, in every component $k$,
\begin{equation}
  \sqrt{\frac{1}{N_k} \sum_{i \in I_k} \bigl(d_i^{\,a} - d_i^{\,b}\bigr)^2} < 0.3\,\sigma_{\pt,k},
  \label{eq:rerun}
\end{equation}
with the comparison points $I_k$ and the scales $\sigma_{\pt,k}$ of Section~\ref{sec:pooled},
computed from all runs. Runs form a group only if every pair in it meets \eqref{eq:rerun}. The
limit $0.3\,\sigma_\pt$ is the one below which ISO 13528 regards the uncertainty of the assigned
value as negligible. What was changed between the runs does not matter, only whether the result
changed: a change that alters it by more than the limit keeps its own vote. Grouping affects only
the spread; every run is still scored on its own.

\textbf{Coverage.} A result contributes to $\sigma_{\pt,i}$ at every point it covers, whether or
not it is complete elsewhere. A point needs at least five results; points with fewer are excluded
from the comparison for all runs.

\subsection{One scale per component}
\label{sec:pooled}

For each component $k$ of a variant, the \emph{comparison points} $I_k$ are the measurement points
that have an experimental value and at least five results; $N_k$ is their number. The
point-wise spreads are pooled into a single scale for each component:
\begin{equation}
  \sigma_{\pt,k} = \sqrt{\frac{1}{N_k} \sum_{i \in I_k} \sigma_{\pt,i}^{\,2}},
  \label{eq:pooled}
\end{equation}
that is, the root mean square of the point-wise spreads. Section~\ref{sec:variances} explains why
the spreads are pooled in this way rather than simply averaged. The surface pressure probes form
one component. All velocity points form one component, across all lines. Each force coefficient
is a single point, so its scale is simply its $\sigma_\pt$.

\begin{figure}[htbp]
  \centering
  \includegraphics[width=0.75\textwidth]{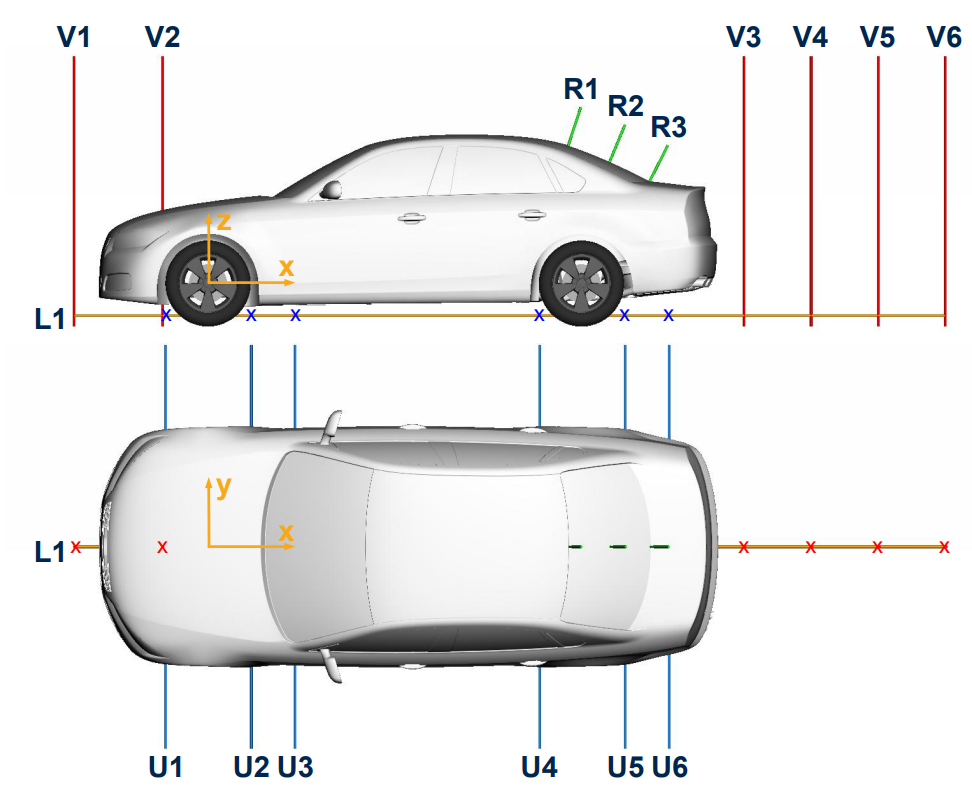}
  \caption{The 16 velocity measurement lines in side view (top) and plan view (bottom): the
    longitudinal line L1, the lateral lines U1--U6, the vertical lines V1--V6 and the rear lines
    R1--R3. 2a and 2b are compared on L1, U2, U3, U5, U6 and V3.}
  \label{fig:lines}
\end{figure}

DD has its own five scales. The configurations 2a and 2b share theirs
(Section~\ref{sec:shared}): the root mean square \eqref{eq:pooled} is taken over the comparison
points of both configurations together,
\begin{equation}
  \sigma_{\pt,k} = \sqrt{\frac{\sum_{i \in I_k^{\mathrm{2a}}} \sigma_{\pt,i}^{\,2}
                               + \sum_{i \in I_k^{\mathrm{2b}}} \sigma_{\pt,i}^{\,2}}
                              {N_k^{\mathrm{2a}} + N_k^{\mathrm{2b}}}},
  \label{eq:shared}
\end{equation}
which is $\sqrt{(\sigma_{\pt,\mathrm{2a}}^{\,2} + \sigma_{\pt,\mathrm{2b}}^{\,2})/2}$ for a force
coefficient. The point-wise spreads $\sigma_{\pt,i}$ remain those of each configuration. For 2a and
2b only the measurements present in both configurations are compared: the $C_p$ probes measured in
both, and the six velocity lines L1, U2, U3, U5, U6 and V3 (Figure~\ref{fig:lines}). The lines U1,
U4, V1, V2 and V5, measured in 2a only, are not used. DD exists only where both configurations were
measured and uses the same probes and lines.

\subsection{Component errors and joint error}
\label{sec:errors}

The error of a run in component $k$ is its root mean square deviation divided by the pooled scale:
\begin{equation}
  e_k = \frac{1}{\sigma_{\pt,k}} \sqrt{\frac{1}{N_k} \sum_{i \in I_k} d_i^{\,2}},
  \label{eq:error}
\end{equation}
where $I_k$ and $N_k$ are the comparison points of the run's own variant and $\sigma_{\pt,k}$ is that
variant's scale: \eqref{eq:pooled} for DD, \eqref{eq:shared} for 2a and 2b. For a force coefficient
the error reduces to $e_k = |d|/\sigma_{\pt,k}$, the magnitude of a $z$ score \eqref{eq:z}. The joint error is the weighted sum
\begin{equation}
  E = \sum_k w_k\, e_k,
\end{equation}
with the default weights of Table~\ref{tab:weights}. Profiles carry 75\,\% of the weight and
forces 25\,\%. The default weights follow the scoring methodology of the AutoCFD5 AI/ML technical
focus group (TFG). They are the main judgement in the metric, and other weightings are defensible.
The workshop's interactive data dashboard will therefore allow users to enter their own weights
and see the effect on the ranking.

\begin{table}[htbp]
  \centering
  \caption{Components and default weights.}
  \label{tab:weights}
  \begin{tabular}{lcc}
    \toprule
    Component & Points per variant & Weight $w_k$ \\
    \midrule
    Surface pressure $C_p$ & about 200 & 0.30 \\
    Velocity $V/V_\infty$, six lines & about 400 & 0.45 \\
    Drag $C_D$ & 1 & 0.15 \\
    Lift $C_L$ & 1 & 0.05 \\
    Pitching moment $C_M$ & 1 & 0.05 \\
    \midrule
    Joint error $E$ & & 1.00 \\
    \bottomrule
  \end{tabular}
\end{table}

\subsection{Completeness}

A run receives a value for component $k$ only if it covers every comparison point in $I_k$
(Section~\ref{sec:pooled}). It receives a joint error only if it has a value for every component.
A run with missing data therefore does not appear on the error--cost diagram, and leaving data out
can never lower a run's error. Component errors of incomplete runs are still reported where they
exist.

\subsection{Freezing the scales}

The joint error depends on the whole field through the scales. To make scores reproducible, the
scales $\sigma_{\pt,k}$ are computed once from all AutoCFD5 Case 2 runs received by 28 October 2026
and then frozen. They are published together with the number of runs behind each. Runs received or
corrected after the cutoff are scored against the frozen scales and do not change them. A future
workshop with the same measurement points can be scored against the same scales, so that progress
between workshops becomes visible.

\section{Design choices}

\subsection{Why pool the spread instead of using a \texorpdfstring{$z$}{z} score at every point?}

The obvious alternative to \eqref{eq:error} is to compute a $z$ score at every point and take the
root mean square of the $z$ scores. That weights every point by $1/\sigma_{\pt,i}^{\,2}$. Points where
the field agrees closely, typically in attached flow, would then dominate, and points where the
field scatters, such as the wake and separated regions, would count for little. Yet those are the
regions where codes and set-ups differ most, and where a ranking should be most sensitive. Pooling
the spread gives every point of a component the same weight, as in a plain root mean square
deviation. Regions of large scatter then count in proportion to the deviations they produce.

The $C_p$ reference shift sharpens the point. All runs agree exactly at probe 107 by construction,
and their scatter is reduced at the probes around it. Point-wise $z$ scores would give those probes
a large weight purely because of where the anchor sits. With pooling they carry no special weight.

\subsection{Why pool the variances rather than average the spreads?}
\label{sec:variances}

The pooled scale \eqref{eq:pooled} is the root mean square of the point-wise spreads, not their
arithmetic mean $\overline{\sigma}$. There are three reasons.

\textbf{Variances add, standard deviations do not.} Combining standard deviations through their
variances is the usual way to pool them, as in the pooled standard deviation of an analysis of
variance.

\textbf{Like is divided by like.} The numerator of \eqref{eq:error} is a root mean square of
deviations, so the scale should be a root mean square of spreads. With the scale
\eqref{eq:pooled} of the run's own variant, \eqref{eq:error} becomes
\[
  e_k = \sqrt{\frac{\sum_{i \in I_k} d_i^{\,2}}{\sum_{i \in I_k} \sigma_{\pt,i}^{\,2}}},
\]
the square root of the run's total squared deviation divided by the field's total variance over the
same points.

\textbf{A typical run scores 1.} If a run's deviation at every point equals the local spread,
$|d_i| = \sigma_{\pt,i}$, then $e_k = 1$ exactly, however much the spread varies from point to
point. The same holds on average for a run drawn at random from a field centred on the experiment,
since then the expected value of $d_i^2$ is $\sigma_{\pt,i}^{\,2}$. With the arithmetic mean, such a
run would score $\sqrt{\overline{\sigma^2}}/\overline{\sigma} > 1$ whenever the spread varies. If
half the points have spread $\sigma$ and half $3\sigma$, for example, it would score 1.12 instead
of 1.

\subsection{Why not take the spread of each run's integrated error?}

Another alternative is to compute one integrated error per run and component, for example the RMS
$C_p$ deviation, and take the spread of these values across runs. That spread measures how much the
runs differ in \emph{quality}, not how far apart their \emph{predictions} are. If all runs had an
RMS deviation of $0.10 \pm 0.01$, every run would score about 10; a field of mixed quality would
score better than a field of uniformly good runs. It would also lose the link to ISO 13528.
Pooling the point-wise spread, as in \eqref{eq:pooled}, keeps one scale per component while
measuring the right quantity.

\subsection{Does the reference probe need special treatment?}
\label{sec:probe}

Only in that it is not compared. After the shift every run matches the experiment at probe 107
exactly, so its deviation and its spread are zero and it says nothing about the run. Left in, it
would still count in $N_k$, and with the shared scale of 2a and 2b its effect would no longer cancel
exactly.

\subsection{Why do 2a and 2b share their scales?}
\label{sec:shared}

2a and 2b are two configurations of the same car, and the workshop compares codes and set-ups, not
configurations. With a scale per configuration, an error of 1 would mean one 2a-field spread in 2a
and one 2b-field spread in 2b. Fewer runs were submitted for 2b, from a partly different set of
participants, so the two units differ. With a shared scale, equal errors mean equal RMS deviations
from the experiment in either configuration. A configuration that is harder to predict then shows
higher errors, which is a finding about the CFD, not an artefact of the scaling.

The spreads are pooled over points, not computed from the 2a and 2b results at a point together:
such a sample would mix the effect of the configuration change into the spread and count runs with
both configurations twice. All runs contribute, not only those with both configurations, since the
shared scale already makes the configurations comparable.

Only measurements present in both configurations are compared. A run's velocity error depends on
where the lines are: a line in the wake, where all codes deviate more, raises it, and the upstream
line L1 lowers it. With a scale per configuration this did not matter, because each scale was taken
over the same lines as the errors it divides and rose and fell with them. A shared scale does not
follow the lines of each configuration. If 2a were scored on its eleven lines and 2b on six, part of
the difference between their errors would come from the choice of lines, not from the CFD. Both are
therefore scored on the six lines they share. The five lines that only 2a has, 272 of its 631
velocity points, are not used.

Every comparison point counts once, so the two configurations carry almost equal weight in the
shared scales: equal for $C_p$ and the forces, 47 and 53\,\% for velocity. On the shared lines, 2b
has more points on L1 than 2a, so L1 makes up 23\,\% of a 2b velocity error and 16\,\% of a 2a one.
Even if a run matched the experiment perfectly on L1, its 2b velocity error would be only about
4\,\% lower than its 2a one for this reason, too little to justify weighting lines rather than
points.

\subsection{Why a robust spread?}

With the ordinary standard deviation, a single failed or misconfigured run would widen the scale
at the affected points and make every other run look better. Discarding such runs would require a
rule for what counts as an outlier, which is itself a judgement. Algorithm A avoids both: it is a
standardised compromise between ignoring a distant result and letting it dominate.

\subsection{Why an error rather than a score?}

A Pareto front can be built from any combination of axis directions: maximising a score and
minimising cost gives the same front as minimising an error and minimising cost. More generally,
any strictly increasing or decreasing transformation of an axis, such as plotting cost on a
logarithmic scale, leaves the front unchanged. We use the joint error because it is zero for
perfect agreement and grows in proportion to the deviations. Unlike a score derived from $R^2$, it
is linear in the deviations within every component, so the weights act as stated over the whole
range and do not compress the differences between good runs.

\section{Interpretation}
\label{sec:interpretation}

A component error $e_k = 1$ means that the run's RMS deviation from the experiment equals the RMS
participant spread for that component. For a single $z$ score, ISO 13528 (clause 9.4) treats
$|z| \le 2$ as acceptable, $2 < |z| < 3$ as a warning signal and $|z| \ge 3$ as an action signal.
Since $E$ is a weighted combination of RMS values, these limits give only a rough sense of
magnitude. The joint error is intended for ranking and for the error--cost diagram, not for pass or
fail verdicts.

For 2a and 2b, which share their scales, a run typical of its field scores 1 only on average over
both configurations. If the 2b field is 20\,\% wider than the 2a field and both have equally many
points, a typical run scores about 0.91 in 2a and 1.09 in 2b: the difference in difficulty shows
instead of being divided out.

The scale is the reproducibility of the field, not an accuracy target. If all participants share a
common deviation from the experiment, for example from the turbulence modelling or from a
difference between the wind tunnel and the simulated set-up, all errors will be large. That is
information, not a defect: it shows that the field agrees more closely with itself than with the
experiment.

A single number hides trade-offs: a run that is good on forces and poor on profiles can have the
same joint error as the reverse. The component errors are therefore reported alongside $E$ and can
be plotted against cost in the same way.

\section{Limitations of a spread taken from the participants}

ISO 13528 permits a $\sigma_\pt$ derived from the participants' results (clause 8.6), but its
introduction (clause 0.2) notes that criteria set independently of the participants, in particular
criteria based on fitness for an end use, generally have advantages. No such criterion exists for
Case 2: there is no agreed accuracy requirement for surface pressure, velocity and forces on this
geometry, and a hand-picked tolerance would reintroduce hidden weights
(Section~\ref{sec:scale}). Freezing the scales moves the method towards the standard's preference.
For any later workshop, the frozen scales are criteria taken from a previous round, the approach
of clause 8.3.

Clause 8.6 also lists disadvantages of a participant-derived $\sigma_\pt$, and clause 5.4 cautions
about small numbers of participants. Table~\ref{tab:limitations} shows how they bear on this
application.

The uncertainty in the scales does not affect rankings \emph{within} a component, because all runs
are divided by the same value. Between components it acts like an uncertainty of similar size in
the effective weights, which is modest compared with the judgement inherent in the weights
themselves. Once frozen, the scales are fixed parts of the metric's definition rather than
estimates that move with each new submission.

\begin{table}[htbp]
  \centering
  \small
  \caption{Limitations named by ISO 13528 and their status here.}
  \label{tab:limitations}
  \begin{tabularx}{\textwidth}{>{\raggedright\arraybackslash}p{0.36\textwidth}
                               >{\raggedright\arraybackslash}X}
    \toprule
    Limitation & Status in AutoCFD5 Case 2 \\
    \midrule
    $\sigma_\pt$ varies from round to round, so scores from different rounds are not comparable.
      & Resolved. The scales are frozen at a cutoff date and reused, so all scores refer to the
        same scales. \\
    \addlinespace
    A scale derived from the field yields a roughly constant share of good scores; generally poor
    or generally good performance is not detected.
      & Not relevant to ranking. The metric orders runs and does not issue verdicts. A field-wide
        deviation from the experiment remains visible in the size of the errors
        (Section~\ref{sec:interpretation}). \\
    \addlinespace
    The scale has no interpretation in terms of fitness for an end use.
      & Accepted. The metric compares runs with one another; it does not certify fitness for
        purpose. \\
    \addlinespace
    With few participants the spread is uncertain, and a small field may be an inappropriate
    comparison group.
      & Mitigated. Around 60 runs were submitted, and every point requires at least five. For a
        single measurand and 60 normally distributed results, a simulation of Algorithm A gives a
        95\,\% range of about $-20$\,\% to $+22$\,\% for the estimated spread. The pooled profile
        scales are steadier. \\
    \bottomrule
  \end{tabularx}
\end{table}

Three points remain open to judgement and are stated for transparency.
\begin{itemize}
  \item \textbf{Runs, not participants.} Each run counts as one result, whereas ISO 13528 assumes
    one result per laboratory. Runs that cannot be told apart count once, so repeating a set-up
    does not move the scale, but different set-ups of one participant count separately. Because
    such runs are correlated, the effective size of the field is smaller than the number of
    results.
  \item \textbf{Composition of the field.} The frozen scales reflect the codes, set-ups and
    participants of AutoCFD5. A future workshop scored against them is compared with the AutoCFD5
    field, which is the intention, but the scales should not be read as universal.
  \item \textbf{Uncertainty of the experiment.} The method treats the wind tunnel values as exact.
    Where the measurement uncertainty $u$ is not small compared with the participants' spread
    (ISO~13528 uses $0.3\,\sigma_\pt$ as the limit), runs are penalised for differences the
    measurement cannot resolve. The standard's $z'$ score (clause 9.5) addresses this; its pooled
    equivalent replaces $\sigma_{\pt,i}^{\,2}$ in \eqref{eq:pooled} and \eqref{eq:shared} by
    $\sigma_{\pt,i}^{\,2} + u_i^2$. It will be adopted once the uncertainties of the wind
    tunnel data are available.
\end{itemize}

\section{Summary of definitions and parameters}

{\small
\begin{tabularx}{\textwidth}{>{\raggedright\arraybackslash}p{0.31\textwidth}
                             >{\raggedright\arraybackslash}X
                             >{\raggedright\arraybackslash}p{0.21\textwidth}}
  \toprule
  Item & Definition or value & Basis \\
  \midrule
  Assigned value & Wind tunnel result & ISO 13528 \\
  Point-wise spread $\sigma_{\pt,i}$ & Algorithm A over all runs at the point & ISO 13528, Annex C \\
  Algorithm A constants & 1.483, 1.5, 1.134 & ISO 13528, Annex C \\
  Results per point & Every valid run, but a participant's runs within 0.3 scales in every
    component count once & Choice; limit ISO 13528 \\
  Minimum results per point & 5 & Choice \\
  Component scale $\sigma_{\pt,k}$ & RMS of $\sigma_{\pt,i}$ over the comparison points $I_k$,
    eq.~\eqref{eq:pooled}; shared by 2a and 2b, eq.~\eqref{eq:shared} & Choice \\
  Compared measurements, 2a and 2b & Those present in both: $C_p$ probes; lines L1, U2, U3, U5, U6,
    V3 & Choice \\
  Velocity scale & One for all lines & Choice \\
  Component error $e_k$ & RMS deviation $/\,\sigma_{\pt,k}$, eq.~\eqref{eq:error} & Choice \\
  Default weights $C_p$ / $V$ / $C_D$ / $C_L$ / $C_M$
    & 0.30 / 0.45 / 0.15 / 0.05 / 0.05; adjustable in the dashboard & AI/ML TFG scoring \\
  $C_p$ reference & Shift to match the experiment at probe 107, which is not compared
    & Case 2 definition \\
  End values of velocity lines & Held over at most 2 experimental points & Choice \\
  Completeness & Every point of every component, for $E$ & Choice \\
  Scales & Frozen from all runs received by 28 October 2026 & Choice \\
  \bottomrule
\end{tabularx}\par
}

\vspace{1em}
\begin{note}
\textbf{Note on the use of AI.} Claude Opus 5.5 (Anthropic) was used to research and document the
approach described here. The authors reviewed the content and are responsible for it.
\end{note}

\end{document}